\documentclass[aps, prd, superscriptaddress, twocolumn,10pt, nofootinbib,preprintnumbers]{revtex4} 

\usepackage[utf8]{inputenc}
\usepackage{graphicx}
\usepackage{dcolumn}
\usepackage{bm}
\usepackage{amsmath}
\usepackage{amsfonts}
\usepackage{amssymb}
\usepackage{wasysym}
\usepackage{float}
\usepackage{ulem}
\begin{document}
\title{Extending the CRESST-II commissioning run limits to lower masses}
\author{Andrew Brown}
\email{a.brown1@physics.ox.ac.uk}
\author{Sam Henry}
\author{Hans Kraus}
\affiliation{Department of Physics, University of Oxford, Keble Road, Oxford OX1 3RH, UK}
\author{Christopher McCabe}
\affiliation{Rudolf Peierls Centre for Theoretical Physics, University of Oxford,\\
1 Keble Road, Oxford OX1 3NP, UK}
\date{\today}

\preprint{OUTP-11-50-P}

\begin{abstract}

Motivated by the recent interest in light WIMPs of mass $\sim\!\!\mathcal{O}(10\,\textup{GeV/c}^2)$, an extension of the elastic, spin-independent WIMP-nucleon cross-section limits resulting from the CRESST-II commissioning run (2007) are presented. Previously, these data were used to set cross-section limits from $1000\,\textup{GeV/c}^2$ down to $\sim\!\!17\,\textup{GeV/c}^2$, using tungsten recoils, in 47.9 kg-days of exposure of calcium tungstate. Here, the overlap of the oxygen and calcium bands with the acceptance region of the commissioning run data set is reconstructed using previously published quenching factors. The resulting elastic WIMP cross-section limits, accounting for the additional exposure of oxygen and calcium, are presented down to $5\,\textup{GeV/c}^2$.

\end{abstract}

\maketitle

\section{Introduction}

Among the outstanding problems in astro-particle physics is to identify the non-baryonic dark matter that makes up $\sim\!80\%$ of the matter in the Universe \cite{Bertone:2004pz}. A well-motivated class of dark matter candidates are Weakly Interacting Massive Particles (WIMPs), which may be directly detected via their elastic scattering with nuclei in terrestrial detectors.

In the CRESST-II commissioning run of 2007 \cite{COMRUN}, limits on elastic, spin-independent, WIMP-nucleon cross-sections for WIMP masses from $1000\,\textup{GeV/c}^{2}$ down to $\sim\!\!17\,\textup{GeV/c}^{2}$ were presented. However, light WIMPs, with a mass \mbox{$\sim\!\!\mathcal{O}(10\,\textup{GeV/c}^2)$} have been suggested as a possible interpretation to the experimental results of DAMA, CoGeNT and CRESST-II (2011) \cite{DAMA, COGENT, NEWCRESST}. At the same time, several other experiments \cite{Ahmed:2009zw, Ahmed:2010wy, XENON, Angle:2011th, CDMSEDELWEISS, Felizardo:2011uw} partially or completely exclude the light WIMP interpretations of \cite{DAMA,COGENT,NEWCRESST}. Therefore, it is of interest to see how data from the commissioning run compare to these results when extended to examine such light WIMP scenarios.

\section{Including Oxygen and Calcium}

The CRESST-II commissioning run data set \cite{COMRUN} consists of 47.9 kg-days exposure of calcium tungstate (CaWO$_4$), taken between the 27$^{\textup{th}}$ of March and the 23$^{\textup{rd}}$ of July, 2007. The data were obtained from two independent modules, labelled by their phonon detector / light detector names: ``Zora/SOS23'' and ``Verena/SOS21'', collecting $23.8$ and $24.1$ kg-days respectively \cite{LANGPAPER}.

\begin{figure}[t!]
\includegraphics[width=0.47\textwidth]{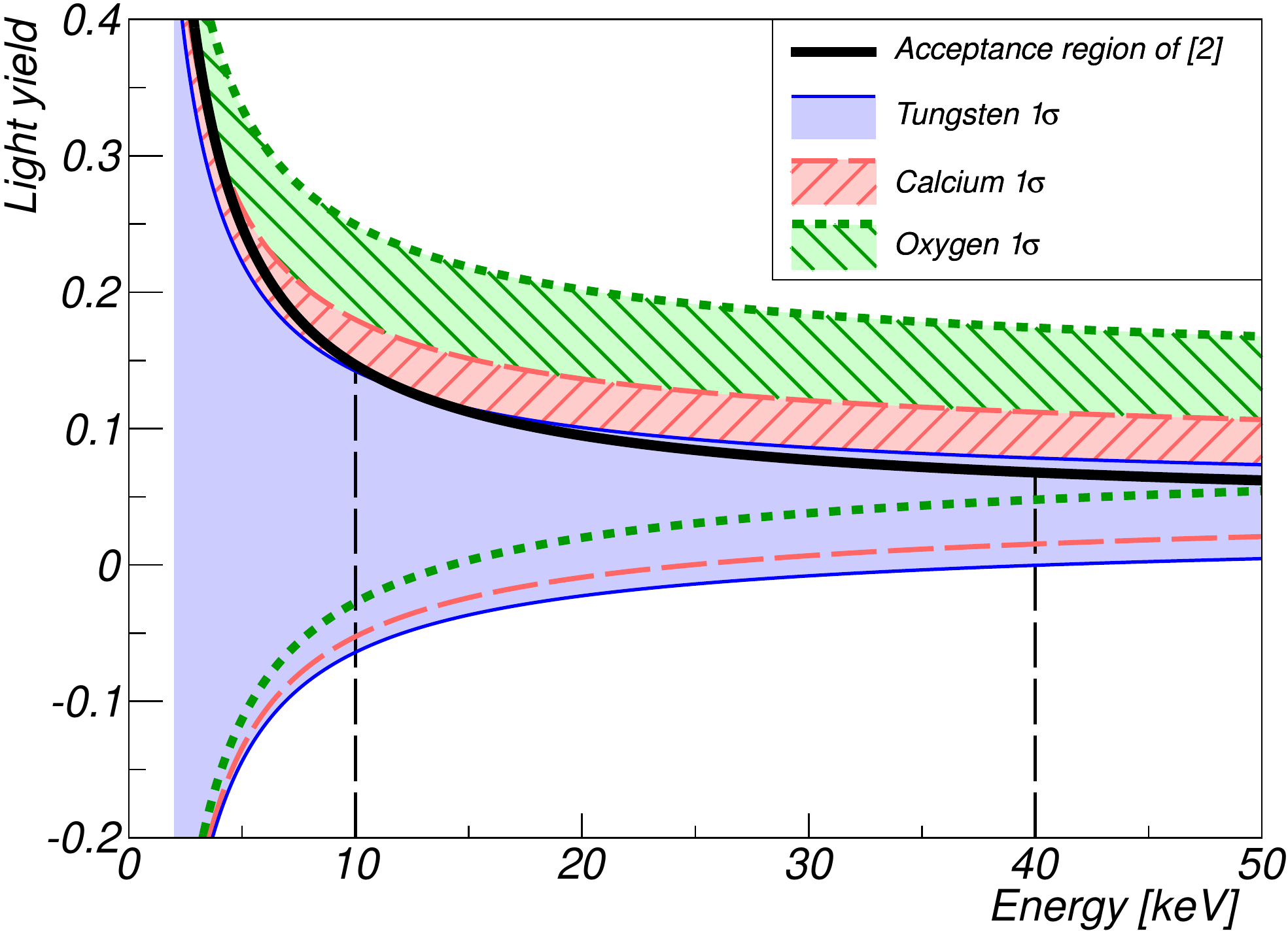}
\caption {Acceptance region and nuclear recoil band diagram for Zora/SOS23. The black line indicates the light yield ($L_{\gamma}/E$) limit for the acceptance region in \cite{COMRUN}, and the vertical dashed lines the acceptance region's lower and upper energy thresholds. $1\,\sigma$ regions for observing recoiling oxygen (green), calcium (red) and tungsten (blue) nuclei are indicated, using light detector resolution parameters in Table \ref{ResTable} and quenching factors from \cite{HUFF}. A similar diagram may be drawn for Verena/SOS21.} 
\label{BandsDiagram}
\end{figure}

In CRESST-II, the phonon detector is used to measure energy and the light detector to distinguish between event types, as different types of interaction in CaWO$_4$ produce different amounts of light, relative to the energy deposited. Gamma and beta interactions, causing electron recoils, produce more light than nuclear recoils. By convention the light produced per unit energy (the `light yield') of electron recoils $L_{\gamma}/E$, is calibrated to those of gamma interactions at $122.06\,\textup{keV}$, which is normalised to a value of one. Nuclear recoils see a reduction in light compared to electron recoils, quantified by a quenching factor, $Q_{i}$, dependent upon the species $i$ of recoiling nucleus. 

The limits of \cite{COMRUN} resulted from assuming all recoils in the acceptance region, the region in which WIMP-nucleon interactions are searched for, were from tungsten alone. However, due to the effects of a finite light detector resolution, recoils from both calcium and oxygen may also be seen within the same acceptance region. This effect is illustrated in Figure \ref{BandsDiagram}. The additional exposure provided by the parts of calcium and oxygen bands that fall within the acceptance region can strengthen cross-section limits for light WIMPs, with mass $\sim\!\!\mathcal{O}(10\,\textup{GeV/c}^2)$. In \cite{COMRUN}, the acceptance region was chosen so that tungsten recoils would have been seen with minimal electron recoil band overlap. In energy, this was between 10 and $40\,\textup{keV}$. In light yield, the upper limit was set so that 90\% of tungsten recoils would occur in the acceptance region. Here, we use this same acceptance region, so that we do not introduce non-blind elements into the analysis.

To consider oxygen and calcium recoils in this region, the fraction of each nuclear species' recoils that fall within the acceptance region must be estimated. This requires several pieces of information. The first is the light detector resolution of the observed electron recoil band, as a function of energy. This resolution is expressed as:
\begin{equation}
\sigma_\gamma^{2}(E)=\sigma_0^{2}+\sigma_1^{2}E+\sigma_2^{2}E^2\,,
\end{equation}	
where $E$ is the energy in the phonon channel. The resolution of the electron recoil band depends on three terms: $\sigma_0$, reflecting electronic noise; $\sigma_1$, related to the Poisson distribution of the expected number of detected photons; and $\sigma_2$, incorporating position dependence and other possible effects seen in CRESST-II light detectors. 

\renewcommand{\arraystretch}{1.25}
\begin{table}[t!]
\begin{center} 
\begin{tabular}{|c|c|c|}
 \hline
  Parameter & Zora/SOS23 & Verena/SOS21 \\ \hline
  $\sigma_{0}\,(\textup{keV})$ & 0.784 & 1.508 \\ 
  $\sigma_{1}\,(\textup{keV}^\frac{1}{2})$ & 1.064 & 0.610 \\
  $\sigma_{2}$ & 0.192 & 0.154 \\  \hline
  $P_0\,(\textup{keV})$ & 0.56 & 0.11 \\ 
  $P_1$ & 0.0040 & 0.0065 \\  \hline
\end{tabular}	
\end{center}
\caption{Resolution values used in this work. $\sigma_{0}$, $\sigma_{1}$ and $\sigma_{2}$ are light detector resolution parameters determined from \cite{COMRUN}, using the method outlined in Appendix \ref{A}. $P_{0}$ and $P_{1}$ are the phonon energy resolution parameters from \cite{LANGTHESIS}.}
\label{ResTable}
\end{table}

An additional piece of information needed is the resolution of each quenched band. At energy $E$, events in the quenched band produce an average amount of light $L_{Q_i}(E)=Q_{i}L_\gamma(E)$. The resolution of a quenched band is assumed to be equal to the resolution of the electron recoil band at energy $E'$, where $L_{Q_i}(E)=L_\gamma(E')$. The exact light detector resolutions used in \cite{COMRUN} are unavailable. However, these resolutions may be obtained by fitting to the acceptance region figures in \cite{COMRUN}, using the method outlined in Appendix \ref{A}. These light detector resolutions are given in Table \ref{ResTable}. Separately, the energy resolution of the phonon detector was modelled in \cite{LANGTHESIS} by $\Delta E = P_{0} +P_{1}E\,,$ with energy resolution parameters also shown in Table \ref{ResTable}.

\begin{figure}[t!]
\includegraphics[width=0.47\textwidth]{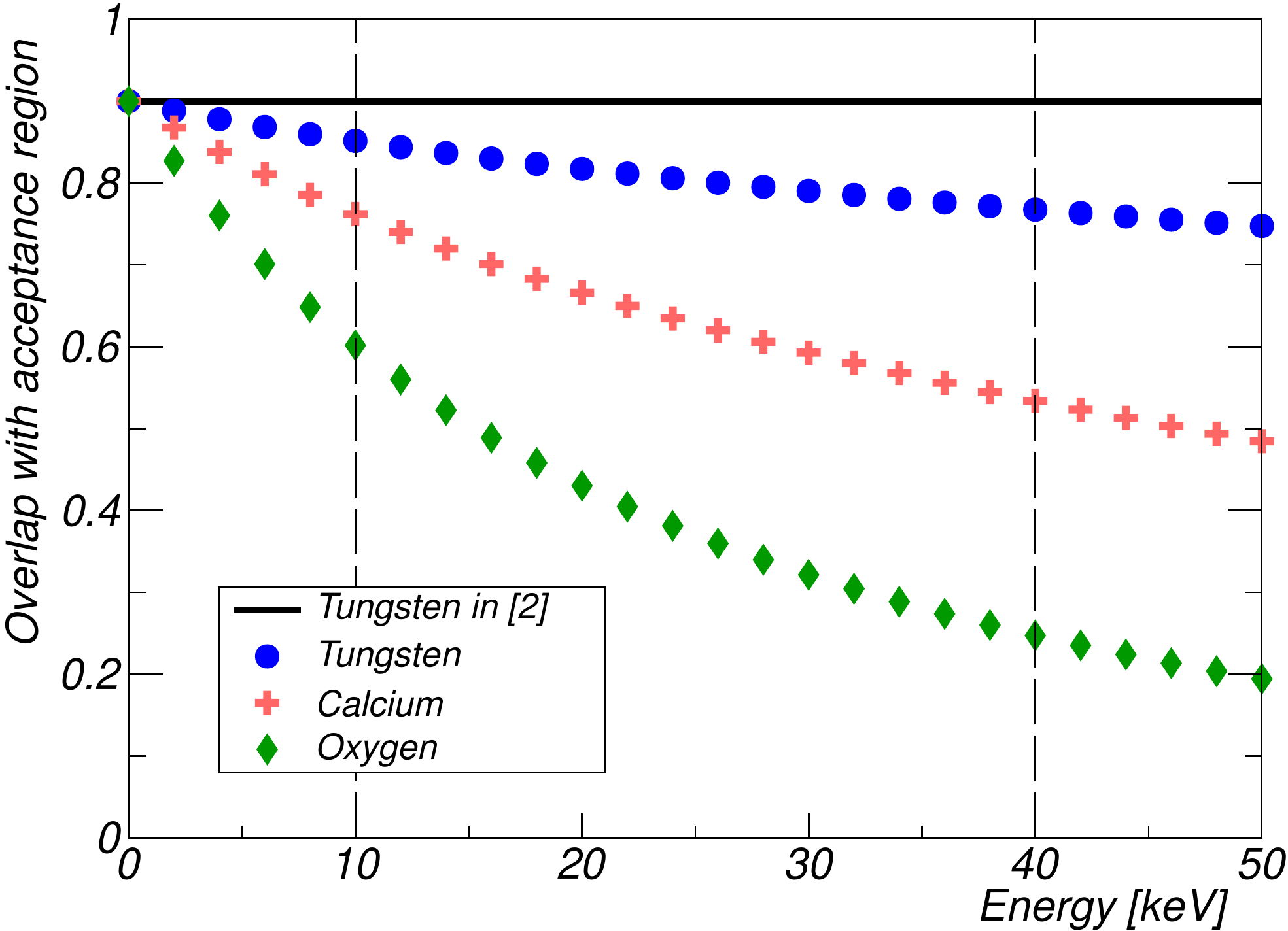}
\caption {Estimates of the fraction of recoils of a given target nucleus that fall within the acceptance region for Zora/SOS23 in \cite{COMRUN} as a function of energy. Oxygen is shown by green diamonds, calcium by red crosses and tungsten by blue circles, with black showing the initial assumption of 90\% of tungsten recoils falling within the acceptance region. Vertical dashed lines at 10 and 40 keV indicate acceptance region limits in recoil energy.} 
\label{OverlapPercent}
\end{figure}

The next piece of information is the quenching factor of each target nucleus. For this, the measurements in \cite{HUFF} are used, with $11.09^{+0.909}_{-0.908}\%$ for oxygen, $6.38^{+0.619}_{-0.653}\%$ for calcium and $3.91^{+0.478}_{-0.430}\%$ for tungsten.\footnote[1]{An oxygen quenching factor of $10.4^{+0.5}_{-0.5}\%$ was used in \cite{NEWCRESST}. Limits calculated using this quenching factor are slightly stronger than those presented here, a result of a larger fraction of oxygen recoils being observable within the acceptance region.} It should be noted that the more recent measurements of light output for tungsten recoils in CaWO$_4$ in \cite{HUFF} are higher than the 2.5\% used in \cite{COMRUN}. This means that the amount of light for tungsten recoils is on average higher than that which was expected in \cite{COMRUN}, causing less than the expected 90\% of all tungsten recoils to fall within the acceptance region, as can be seen in Figure \ref{OverlapPercent} for Zora/SOS23.

One last piece of information would be required for a complete description of detected light. This is the small deviation of observed light in the electron recoil band from the normalisation of one unit of light per unit energy. Two effects can cause this deviation: the dependence of light yield on energy in inorganic scintillators \cite{INORGSCIN}, and an overall calibration error. Such adjustments as used in the analysis of \cite{COMRUN} are unavailable, although it is stated in \cite{COMRUN} that the light yield is always near the normalisation of one. Here we use the approximation that the mean electron recoil light yield is one everywhere. In an independent analysis of the commissioning run data [14], the electron recoil band behaviour and light detector resolution were parameterised, with results repeated in Appendix \ref{B}. As a check on our results, we also considered the resulting WIMP cross-section limits with these parameters. They are consistent with those presented here to within a few percent.

With these pieces of information, the fraction of recoiling nuclei from each constituent of CaWO$_{4}$ that falls within the acceptance region of \cite{COMRUN} can be estimated, as shown in Figure \ref{OverlapPercent} for Zora/SOS23. With these fractions, the interaction rate of WIMPs with oxygen and calcium in the commissioning run acceptance region may now be calculated. Here we follow a method analogous to that in \cite{COMRUN}. The elastic, spin-independent WIMP-nucleon interaction rates are calculated following \cite{DONATO}, using the Helm form factor parameterisation as suggested in \cite{LEWINSMITH}. The total rate expected from all target nuclei is then:
\begin{equation}
\frac{dR_{\text{Tot}}}{dE}=A_{W}\frac{dR_{W}}{dE}+A_{Ca}\frac{dR_{Ca}}{dE}+A_{O}\frac{dR_{O}}{dE}\,,
\end{equation}                              				
for the fractions $A_{i}$ of each species' nuclear recoils that may be seen in the acceptance region. This rate is convolved with the observed phonon energy resolution, $\Delta E$, as described in \cite{LEWINSMITH}.
\section{Results} 

\begin{figure}[t!]
\includegraphics[width=0.47\textwidth]{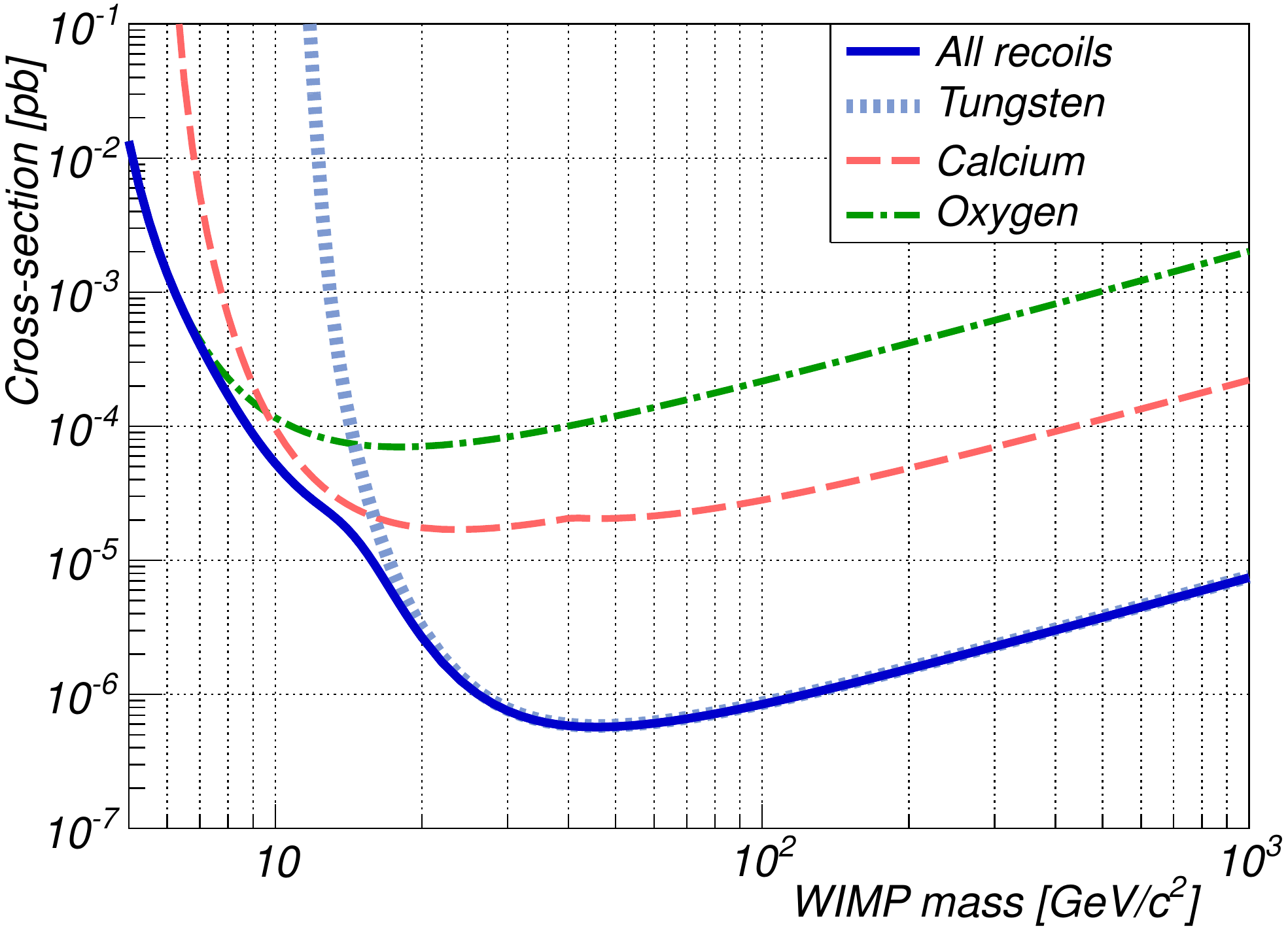}
\caption{90\% confidence limits on elastic, spin-independent, WIMP-nucleon cross-sections from data in \cite{COMRUN}, considering all possible nuclear recoils within the acceptance region. The green dashed-dotted, red dashed and light-blue dotted lines result from considering WIMP interactions with oxygen, calcium and tungsten individually, and the solid blue line the total rate. The WIMP halo properties used are $\rho_{\text{DM}}=0.3\,\textup{GeV/}\textup{cm$^3$}$, $v_{\text{esc}} = 544\,\textup{km/s}$, $v_{0} = 220\,\textup{km/s}$ and $v_{\text{sun}} = 232\,\textup{km/s}$. Resolutions from Table \ref{ResTable} and quenching factors from \cite{HUFF} were used to derive these limits.}
\label{LimitPlot}
\end{figure}

\begin{figure}[t!]
\includegraphics[width=0.47\textwidth]{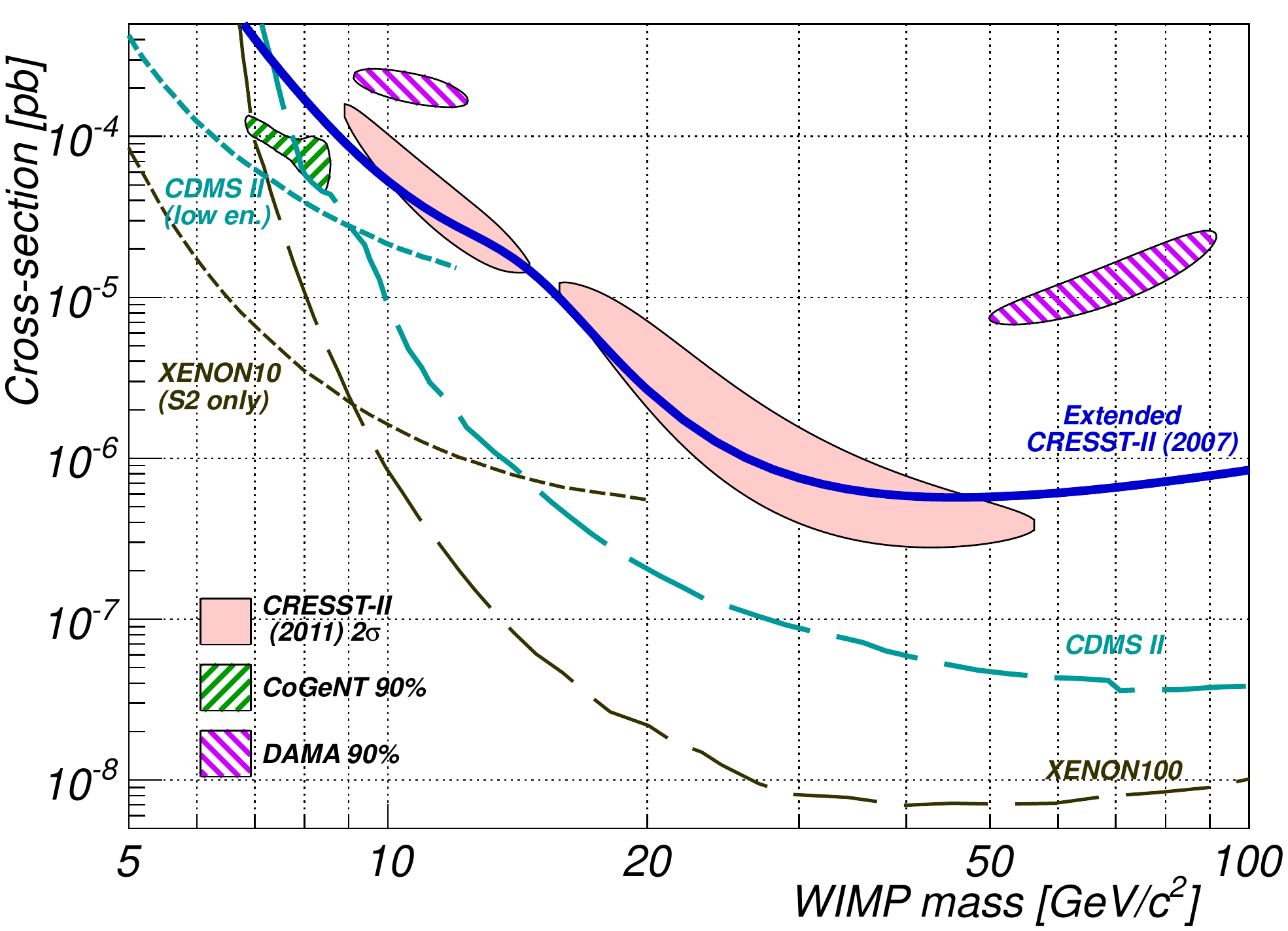}
\caption{The combined 90\% confidence limit on the elastic, spin-independent WIMP-nucleon cross-section from extending the analysis of commissioning run data to lower WIMP masses (solid blue). For comparison, the favoured regions from DAMA, derived from \cite{DAMA}, CoGeNT~\cite{COGENT} and CRESST-II (2011) \cite{NEWCRESST} are shown. Also shown are WIMP cross-section limits from, CDMS II~\cite{Ahmed:2009zw}, CDMS II (low energy)~\cite{Ahmed:2010wy}, XENON100~\cite{XENON} and XENON10~(S2 only)~\cite{Angle:2011th}.}
  
\label{CompareLimits}
\end{figure}

Three events are observed in \cite{COMRUN}, at $16.89\,\textup{keV}$, $18.03\,\textup{keV}$ and $33.09\,\textup{keV}$. The Maximum Gap method \cite{YELLIN} is used to calculate the resulting elastic, spin-independent, WIMP-nucleon cross-section limits. The results are shown in Figure \ref{LimitPlot}. The extended 90\% confidence limit improves sensitivity to low mass WIMPs with commissioning run data. Interactions with WIMPs heavier than $\sim\!17\,\textup{GeV/c}^2$ in the acceptance region are dominated by tungsten recoils, and the cross-section limit is well modelled by considering interactions from tungsten alone. At lower masses, calcium, then oxygen recoils become dominant, such that below $\sim\!7\,\textup{GeV/c}^2$, nearly all WIMP-nucleon interactions in the acceptance region are with oxygen nuclei. Considering all possible nuclear recoils then provides a significant strengthening of cross-section limits at low masses compared to tungsten alone. 

In Figure \ref{CompareLimits}, a comparison of the combined limit is made to the elastic WIMP interpretation of other experiments \cite{DAMA, COGENT, NEWCRESST, XENON,Angle:2011th, Ahmed:2009zw, Ahmed:2010wy}. The CoGeNT, CRESST-II (2011) and DAMA results were already in tension with the results of XENON100, XENON10 (S2 only), CDMS II, and CDMS II (low energy). The extended CRESST-II commissioning run limits introduce further mild tension with DAMA and CRESST-II (2011).

As the commissioning run and CRESST-II (2011) results are with the same target nuclei with similar energy thresholds, it is difficult to reduce this mild tension by choosing different astrophysical parameters or particle physics models. However, it should be noted that the CRESST-II commissioning run and CRESST-II (2011) run do not use the same acceptance region definitions. In this work, we have used the acceptance region defined in the original commissioning run analysis. While this ensures that we have not introduced non-blind elements into the analysis, this region has not been optimised for light mass WIMP discovery. An additional difference between runs is the design of clamps in direct contact with the target crystals, which as noted in \cite{NEWCRESST} introduced additional backgrounds into the CRESST-II (2011) data set. Since the commissioning run live-time is much smaller than in the CRESST-II (2011) run, repeating the commissioning run experimental conditions for a longer period would allow stronger conclusions to be drawn.

\section{Conclusions}

The WIMP cross-section limits for the 47.9 kg-days exposure of CaWO$_4$ in the CRESST-II commissioning run \cite{COMRUN} have been extended down to a WIMP mass of $5\,\textup{GeV/c}^2$. Our analysis has accounted for possible oxygen and calcium recoils within the commissioning run acceptance region, using light and phonon detector resolutions in Table~\ref{ResTable} and quenching factors from \cite{HUFF}. The improvement of cross-section limits at light masses occurs because recoiling oxygen and calcium nuclei dominate over tungsten recoiling nuclei for light WIMPs. Extending the commissioning run limits results in mild tension with the recent CRESST-II \cite{NEWCRESST} and DAMA \cite{DAMA} results.

\section*{Acknowledgments}
We wish to thank Franz Pr\"{o}bst for providing the favoured region contours of recent CRESST-II results, Jens Schmaler for numerous helpful comments on this work, and Felix Kahlhoefer for useful discussions. We also wish to acknowledge the Science and Technology Facilities Council, UK who funded this work.

\appendix

\section{Reconstructing light detector resolutions}

\renewcommand{\arraystretch}{1.25}
\label{A}
To estimate light detector resolutions, oxygen and tungsten acceptance regions were modelled by:
\begin{equation}
\text{Acc}(Q_{i},E)=\frac{Q_{i}L_{\gamma}(E) +N_{\text{sig}}\sigma_{Q_{i}}(E)}{E}\,,
\end{equation}
where $Q_{i}$ is the quenching factor of the considered nucleus, and $N_{\text{sig}}\!\approx\!1.28$ is the number of standard deviations allowing 90\% of quenched recoils to be seen in the acceptance region. These equations were fitted simultaneously to the tungsten and oxygen nuclear recoil acceptance regions in Figure 8 of \cite{COMRUN}. The tungsten quenching factor is set at 2.5\% and $L_{\gamma}/E$  is taken to be one. The oxygen quenching factor, given as $\sim\!11.1\%$ in the text of \cite{COMRUN} is allowed to vary in the fit. A value of 11.0\% is found from fitting to both modules' acceptance regions.

\section{Alternative light detector resolutions}
\label{B}

In \cite{LANGTHESIS}, the electron recoil behaviour was modelled by:
\vspace{-3 mm}
\begin{equation}
L_\gamma(E)=\frac{l_{1}E}{1+e^{-l_{e}E}}\,, 
\vspace{-3 mm}
\end{equation}
\linebreak
giving electron recoil band behaviour parameters and light detector resolutions for Zora/SOS23:

\vspace{2 mm}
\begin{center}
\begin{tabular}{|c|c|}
 \hline
  Parameter & Zora/SOS23 \\ \hline
  $l_1$ & $1.068\pm0.003$  \\ 
  $l_e\,(\textup{keV$^{-1}$})$ & $0.180\pm0.007$  \\  
  $\sigma_{0}\,(\textup{keV})$ & $1.1\pm0.3$ \\ 
  $\sigma_{1}\,(\textup{keV}^\frac{1}{2})$  & $0.46\pm0.03$ \\
  $\sigma_{2}$ & $0.178\pm0.004$ \\  \hline
\end{tabular}	
\end{center}

Light detector resolution for Verena/SOS21 was modelled by three time-separated noise regions with differing light detector resolutions, labelled “High”, “Medium” and “Low” noise regions. The electron recoil band behaviour and light detector resolution parameters were:

\begin{center}
\begin{tabular}{|c|c|c|c|}
 \hline
  Parameter  & \multicolumn{3}{c|}{Verena/SOS21} \\ \cline{2-4}
  & \,\,High\,\, & \,\,Low\,\, & Medium \\ \hline
  $l_1$ & $1.021\pm0.005$ & $1.035\pm0.003$ & $1.036\pm0.002$ \\ 
  $l_e\,(\textup{keV$^{-1}$})$ & $0.169\pm0.001$ & $0.171\pm0.008$ & $0.22\pm0.01$ \\  
  $\sigma_{0}\,(\textup{keV})$ & $3.5\pm0.6$ & $1.18\pm0.37$ & $1.26\pm0.13$\\ 
  $\sigma_{1}\,(\textup{keV}^\frac{1}{2})$ & $1.00\pm0.15$ & $0.72\pm0.07$ & $0.76\pm0.04$ \\
  $\sigma_{2}$ & $0.03\pm0.08$ & $0.09\pm0.01$ & $0.106\pm0.006$ \\  \hline
\end{tabular}	
\end{center}

The live-time distribution between regions is taken from Figure 9.22 of \cite{LANGTHESIS} at 20\%, 32\% and 48\% respectively. However, when using these light detector resolutions, it should be noted that there are some differences between the exact cuts and data selection between \cite{LANGTHESIS} (47.5 kg-days) and \cite{COMRUN} (47.9 kg-days).

\bibliography{ref}
\bibliographystyle{Arxiv}

\end{document}